\documentclass[a4paper,11pt]{article}
\usepackage{jheppub}
\usepackage{CJK}
\usepackage{tikz}
\usepackage{float,graphbox,makecell,tablefootnote}
\usepackage{longtable}
\usepackage{multirow}
\usepackage{booktabs}
\allowdisplaybreaks[4]
\usepackage{diagbox}
\usepackage{alltt}

\definecolor{darkred}{RGB}{173,34,48}
\definecolor{lightgreen}{rgb}{0.56, 0.69, 0.19}
\definecolor{lightblue}{rgb}{0.36, 0.51, 0.71}
\definecolor{lightyellow}{rgb}{0.88, 0.61, 0.14}
\definecolor{darkgreen}{rgb}{0.6, 0.6, 0.35}
\definecolor{lightred}{rgb}{0.99, 0.36, 0.02}
\definecolor{box1}{rgb}{0.46, 0.6, 0.45}
\definecolor{box2}{rgb}{0.62, 0.56, 0.43}
\definecolor{box3}{rgb}{0.72, 0.65, 0.17}

\usepackage[hang]{footmisc}
\makeatletter
\ifFN@para
\else
  \long\def\@makefntext#1{%
    \ifFN@hangfoot
      \bgroup
      \setbox\@tempboxa\hbox{%
        \ifdim\footnotemargin>0pt
          \hb@xt@\footnotemargin{\@makefnmark\hss}%
        \else
          \@makefnmark\hskip-\footnotemargin      
        \fi
      }%
      \leftmargin\wd\@tempboxa
      \rightmargin\z@
      \linewidth \columnwidth
      \advance \linewidth -\leftmargin
      \parshape \@ne \leftmargin \linewidth
      \footnotesize
      \@setpar{{\@@par}}%
      \leavevmode
      \llap{\box\@tempboxa}%
      \parskip\hangfootparskip\relax
      \parindent\hangfootparindent\relax
    \else
      \parindent1em
      \noindent
      \ifdim\footnotemargin>\z@
        \hb@xt@ \footnotemargin{\hss\@makefnmark}%
      \else
        \ifdim\footnotemargin=\z@
          \llap{\@makefnmark}%
        \else
          \llap{\hb@xt@ -\footnotemargin{\@makefnmark\hss}}%
        \fi
      \fi
    \fi
    \footnotelayout#1%
    \ifFN@hangfoot
      \par\egroup
    \fi
  }
\fi
\makeatother

\usepackage{slashed}

\usepackage{tikz}

\usepackage{graphicx}
\usepackage{dcolumn}
\usepackage{bm}
\usepackage{hyperref}
\usepackage{algorithm}        
\usepackage{algpseudocode}    
\usepackage{tikz}
\usetikzlibrary{shapes.misc, positioning}

\newcommand{\rmd}{{\rm d}}
\newcommand{\ii}{\text{i}}
\newcommand{\ee}{\text{e}}

\begin{document}

\begin{CJK*}{UTF8}{}
\CJKfamily{gbsn}

\title{Notes on Diagrammatic Coaction for Cosmological Wavefunction Coefficients: A Two-Site Prelude}

\author[a]{Yuhan Fu (付雨涵),}\emailAdd{yuhanfu1@link.cuhk.edu.cn}
\author[b,c]{Jiahao Liu (刘家昊)}\emailAdd{liujiahao@itp.ac.cn}

\affiliation[a]{School of Data Science, The Chinese University of Hong Kong, Shenzhen, Guangdong 518172, China}
\affiliation[b]{Institute of Theoretical Physics, Chinese Academy of Sciences, Beijing 100190, China}
\affiliation[c]{School of Physical Sciences, University of Chinese Academy of Sciences, No.19A Yuquan Road, Beijing 100049, China}

\abstract{We study the coaction of cosmological wavefunction coefficients of conformally coupled scalars in FRW background of a two-site example, which turns out to have an elegant diagrammatic interpretation. We show how the coaction acts on the twisted integrals for wavefunction coefficients, decomposing them into contributions associated with subtopologies and cuts, with the subtopologies admitting an interpretation as time-ordered integrals. This provides a clear interpretation of their analytic structure and suggests a broader applicability to more general cosmological diagrams.}

\maketitle
\end{CJK*}

\section{Introduction}
In the so-called Cosmological Collider program~\cite{Chen:2009we,Chen:2009zp,Chen:2012ge,Noumi:2012vr,Arkani-Hamed:2015bza,Chen:2016hrz,Chen:2016nrs,Chen:2016uwp,Lee:2016vti,An:2017hlx}, relevant observables are correlation functions of primordial fluctuations, also known as the \textit{cosmological correlators}, and the corresponding \textit{wavefunction coefficients} of the universe. In the context of inflation~\cite{Guth:1980zm, Linde:1981mu, Albrecht:1982wi, Baumann:2009ds}, these correlation functions are calculated on the ``boundary'' of an approximate de Sitter (dS) spacetime, which can be interpreted as scattering amplitudes in dS spacetime. Thus, cosmological correlators serve a dual purpose: they illuminate primordial cosmology while also offering insights into quantum field theory in curved spacetime~\cite{Anninos:2014lwa,Chen:2017ryl,Sleight:2021plv,Stefanyszyn:2023qov,Penedones:2023uqc,Loparco:2023rug,Loparco:2023akg,Marolf:2010nz,Cespedes:2023aal}. Similar to flat-space QFT, a perturbative expansion via Feynman diagrams remains applicable; however, the absence of time-translation invariance introduces additional complexities, such as non-trivial time integrals and more intricate propagator structures~\cite{Maldacena:2002vr,Weinberg:2005vy,Chen:2009we,Chen:2009zp,Chen:2012ge,Noumi:2012vr}.

In recent years, an interesting toy model involving conformal scalar fields with time-dependent self-interactions in the power-law FRW universe has been extensively studied. This model strikes a balance between simplicity---allowing explicit evaluation of higher-order Feynman diagrams---and richness, enabling the study of the aforementioned complexities. Novel techniques have been developed, including partial Mellin-Barnes representation~\cite{Matsubara-Heo:2023ylc,berkesch2014euler}, family-tree decomposition~\cite{Fan:2024iek,fan2025anatomy}, differential equations~\cite{Arkani-Hamed:2023bsv, Arkani-Hamed:2023kig,baumann2025geometry,he2025differential,chen2025multivariate,He:2024olr}, generalization of Ramanujan's master theorem (method of brackets)~\cite{Raman:2025tsg} and cluster bootstrap~\cite{Capuano:2026pgq,Paranjape:2026htn,Mazloumi:2025pmx}. Furthermore, well-established methods from flat-space amplitude computations, notably in $\mathcal{N}=4$ SYM theory, have been adapted to this cosmological context. Remarkably, elegant geometric structures, particularly from the perspective of \textit{positive geometry}, have been constantly emerging after the introduction of cosmological polytope~\cite{Arkani-Hamed:2017fdk,Arkani-Hamed:2018bjr,Benincasa:2019vqr,Benincasa:2018ssx,Benincasa:2020aoj,Juhnke-Kubitzke:2023nrj,Benincasa:2024leu}. This geometric perspective has since proliferated, giving rise to concepts such as kinematic flow for trees and loops~\cite{Arkani-Hamed:2023kig,Arkani-Hamed:2023bsv,Hang:2024xas,Baumann:2024mvm} (the canonical differential equations (CDE) in a broader sense~\cite{Henn:2013pwa,Capuano:2025ehm}), the Cosmohedra~\cite{Arkani-Hamed:2024jbp}.

The integrated cosmological correlators of conformally coupled scalars in FRW spacetime, when expanded around the FRW twist parameter $\epsilon=0$, yield transcendental functions at each order, namely multiple polylogarithms (MPLs), which are ubiquitous in the study of Feynman integrals.  Considerable efforts have been devoted to understanding the algebraic properties of MPL-type Feynman integrals, including Landau analysis, symbology and (diagrammatic) coaction. Landau analysis reveals the singularity structure of the integrals, symbology provides the underlying symbol alphabet, and coaction captures the transcendental constants not fully encoded in the symbol. The concept of coaction for Feynman integrals was first introduced in~\cite{Duhr:2012fh} and later extended to resummed results in the $\epsilon$-expansion, receiving a diagrammatic interpretation through a series of subsequent works~\cite{Abreu:2014cla,Abreu:2015zaa,Abreu:2017ptx,Abreu:2017enx,Abreu:2017mtm,Abreu:2021vhb}.

Meanwhile, the recently introduced approach of twisted cohomology and intersection theory~\cite{Caron-Huot:2021iev,Caron-Huot:2021xqj,aomoto2011theory, yoshida2013hypergeometric, Matsubara-Heo:2023ylc,Frellesvig:2019uqt, Mizera:2019vvs,matsumoto1998kforms} has inspired a series of applications to cosmological correlators~\cite{de2024physicalbasiscosmologicalcorrelators,De:2023xue}. After converting the nested time integrals appearing in cosmological correlators to energy integrals, the resulting twisted integrals feature integrands whose singularities occur exclusively on hyperplanes. Coactions for such integrals have already been explored in~\cite{Abreu:2019wzk}. It is therefore natural to explore whether the coaction structure can be generalized to cosmological correlators and to ask whether a diagrammatic interpretation also exists in this setting. As we will demonstrate, the notion of a \textit{diagrammatic coaction for cosmological correlators} admits a natural and elegant formulation.

This paper is organized as follows. In Section \ref{sec:ii}, we review cosmological wavefunction coefficients in FRW backgrounds and introduce the two-site model. In Section \ref{sec:iii}, we review the diagrammatic coaction and illustrate its action on explicit examples. In Section \ref{sec:iv}, we explicitly calculate the coaction for two-site chain and give a diagrammatic interpretation. We conclude in Section \ref{sec:v} with a discussion of implications and future directions. An appendix is included to show the consistency with the known result and check the coassociativity of the coaction.

\section{Review of Cosmological Correlators}\label{sec:ii}

The self-interacting conformal scalar theory in ($d+1$)-dimensional power-law FRW background is described by the following action
\begin{equation}
    S\left[\phi\right] = - \int \rmd^d x \, \rmd \tau \sqrt{-g} \left( \frac{1}{2} (\partial \phi)^2 + \frac{1}{2} \xi R \phi^2 + \sum_{k >2}\frac{\tilde{\lambda}_k}{k!} \phi^k \right),
\end{equation}
where the conformally flat FRW metric is
\begin{equation}
    \rmd ^2 s = a(\tau)^2 (-\rmd \tau ^2 + \rmd x^2) = \left(\frac{\tau}{\tau_0}\right)^{-2 \epsilon}(-\rmd \tau ^2 + \rmd x^2).
\end{equation}
We are interested in the case where $\xi = (d-1)/(4d)$. After a rescaling $\phi \rightarrow a(\tau)^{\frac{1-d}{2}} \varphi $, the above action is equivalent to a massless scalar theory in flat spacetime
\begin{equation}
    S\left[\varphi\right] = -\int \rmd^d x \, \rmd \tau \left( \frac{1}{2} (\partial \varphi)^2 + \sum_{k>2} \frac{\lambda_k(\tau)}{k!} \varphi^k  \right),
\end{equation}
with time-dependent coupling $\lambda_k(\tau) = \tilde{\lambda}_k (-\tau)^{q_k-1} $, and $q_k=\left( \frac{k(d-1)}{2}-d-1 \right) \epsilon$. Note that when $d=k=3$, we have $q_k=-\epsilon$. This special case ($\phi^3$ theory in $(3+1)$-dimensional spacetime) has been studied extensively over the last couple of years~\cite{Arkani-Hamed:2017fdk,Arkani-Hamed:2018bjr,Benincasa:2019vqr,Benincasa:2018ssx,Benincasa:2020aoj,Juhnke-Kubitzke:2023nrj,Benincasa:2024leu}. In this paper, we will focus on the wavefunction coefficients in this particular situation.

The wavefunction is defined as
\begin{equation}
   \Phi[\hat\varphi]=\int_{\varphi(\tau=-\infty)=|0\rangle}^{\varphi(\tau=0)=|\hat\varphi\rangle}\mathcal{D}\,\varphi \, \ee^{\ii \,  S[\varphi]}
\end{equation}
and can be expanded perturbatively (in momentum space) as
\begin{equation}
   \Phi[\hat\varphi]\equiv \exp{\left[-\ii\, \sum_{n=2}^\infty\frac{1}{n!}\int\prod_{i=1}^n\left(\frac{{\rm d}^d\mathbf{k}_i}{(2\pi)^d}\hat\varphi(\mathbf{k}_i)\right)\hat{\psi}_n(\mathbf{k}_1,\cdots,\mathbf{k}_n)\ \delta^{(d)}\left(\sum_i\mathbf{k}_i\right)\right]},
\end{equation}
where the $n$-site wavefunction coefficients $ \hat{\psi}_n(\mathbf{k}_1,\cdots,\mathbf{k}_n) $ can be calculated by ``Feynman diagrams". And the ``Feynman rules" involve two kinds of propagators: (i) the bulk-to-boundary propagator (between a bulk point at $\tau_v$ and the boundary $\tau=0$, carrying energy $X = |\mathbf{k}|$)
\begin{equation}
    B(X,\tau_v)=\ee^{\ii X\tau_v}.
\end{equation}
If multiple bulk-to-boundary propagators attach to the same vertex $v$, we will have
\begin{equation}
\raisebox{-29pt}{
\begin{tikzpicture}[line width=1. pt, scale=2]
\draw[lightgray, line width=1.pt] (0,0) -- (-0.25,0.55);
\draw[lightgray, line width=1.pt] (0,0) -- (0.25,0.55);
\draw[lightgray, line width=2.pt] (-0.5,0.55) -- (0.5,0.55);
\draw[fill=black] (0,0) circle (.03cm);
\node[scale=1] at (0,-.15) {$v$};
\node[lightgray,scale=1] at (0,0.35) {$...$};
\node[lightgray,scale=.7] at (-0.2,0.2) {$1$};
\node[lightgray,scale=.7] at (0.2,0.2) {$n$};
\end{tikzpicture}
}
\sim \prod_{1 \leq i \leq n} \ee^{\ii X_i \tau_v} = \ee^{\ii X_v \tau_v }, \quad X_v = \sum_{1 \leq i \leq n} X_{i}.
\end{equation}
Thus, it suffices to consider the energy $X_v = \sum_{1 \leq i \leq n} X_{i} $ of this vertex, and omit the bulk-to-boundary details for simplicity. (ii) the bulk-to-bulk propagator (for an edge $e$ with vertices $v_1$ and $v_2$ and carrying energy $Y_e = |\mathbf{K}_e|$, where $ \mathbf{K}_e $ is a sum over some external momenta $\mathbf{k}_i$)
\begin{equation}
\begin{aligned}
    G_e(Y_e,\tau_{v_1},\tau_{v_2}){=}\frac{1}{2Y_e} \left[ \ee^{+\ii \, Y_e(\tau_{v_1}{-}\tau_{v_2})}\theta(\tau_{v_2}{-}\tau_{v_1}) {+} \ee^{-\ii \, Y_e(\tau_{v_1}{-}\tau_{v_2})}\theta(\tau_{v_1}{-}\tau_{v_2}) {-}\ee^{+\ii \, Y_e(\tau_{v_1}+\tau_{v_2})} \right]\!.
\end{aligned}
\end{equation}
We introduce the time-order decomposition of the propagator as in \cite{He:2024olr}:
\begin{equation}\label{eq:timedecomp}
	\begin{aligned}
	\raisebox{-1em}{\begin{tikzpicture}
		\coordinate (X2) at (1,0);
		\coordinate (X3) at (2,0);
		\node[below] at (X2) {\small{$1$}};
		\node[below] at (X3) {\small{$2$}};
        \node at (1.5,0.15) {\small{$e$}};
		\draw[line width=1.5pt] (X2)--(X3);
		\path[fill=black] (X2) circle[radius=0.1];
		\path[fill=black] (X3) circle[radius=0.1];
\end{tikzpicture}} 
 = & \raisebox{-1em}{\begin{tikzpicture}
				\coordinate (X2) at (1,0);
				\coordinate (X3) at (2,0);;
				\node[below] at (X2) {\small{$1$}};
				\node[below] at (X3) {\small{$2$}};
				\draw[line width=1.5pt,->] (X2)--(1.6,0);
				\draw[line width=1.5pt] (1.5,0)--(X3);
				\path[fill=black] (X2) circle[radius=0.1];
				\path[fill=black] (X3) circle[radius=0.1];
		\end{tikzpicture}} + 
		\raisebox{-1em}{\begin{tikzpicture}
		\coordinate (X2) at (1,0);
		\coordinate (X3) at (2,0);;
		\node[below] at (X2) {\small{$1$}};
		\node[below] at (X3) {\small{$2$}};
		\draw[line width=1.5pt,->] (X3)--(1.4,0);
		\draw[line width=1.5pt] (1.5,0)--(X2);
		\path[fill=black] (X2) circle[radius=0.1];
		\path[fill=black] (X3) circle[radius=0.1];
	\end{tikzpicture}} - 
	\raisebox{-1em}{\begin{tikzpicture}
	\coordinate (X2) at (1,0);
	\coordinate (X3) at (2,0);
	\node[below] at (X2) {\small{$1$}};
	\node[below] at (X3) {\small{$2$}};
	\draw[line width=1.5pt,dashed] (X2)--(X3);
	\path[fill=black] (X2) circle[radius=0.1];
	\path[fill=black] (X3) circle[radius=0.1];
\end{tikzpicture}},
	\end{aligned}
\end{equation}
where an arrow pointing from $i$ to $j$ encodes the time-order $\theta(\tau_{j}-\tau_i)$ and dashed line means there is no time-order constraint. For a given $n$-site $L$-loop diagram $\mathcal{G}$ with edges $\mathcal{E}$, the wavefunction coefficient is written as
\begin{equation}
\hat{\psi}_{n}^{(L)}(\mathbf{k}_1,\cdots,\mathbf{k}_n;\mathcal{E})=\int_{-\infty}^0\prod_{v=1}^n{\rm d}\tau_v\ \ii\,\lambda_v(\tau_v) B(X_v,\tau_v)\prod_{e\in\mathcal{E}}G_e(Y_e,\tau_{v_1},\tau_{v_2}).
\end{equation}
There should have been integration over spatial loop momenta for loop-level wavefunction coefficients. In this paper, we focus only on the tree-level integrals and the ``integrand" (w.r.t loop momenta integration) for loop-level integrals. Then we will always suppress the extra integration over spatial loop momenta.

Moreover, we would like to substitute the power of the conformal time, appearing in the time-dependent coupling $\lambda_k(\tau) \propto (-\tau)^{-\epsilon-1} $, by the following integral
\begin{equation}\label{eq:timepowerrep}
    (-\tau)^{-\epsilon-1} = \frac{\ii^{1+\epsilon}}{\Gamma(1+\epsilon)} \int_0^\infty \rmd x \, \ee^{\ii x \tau} x^\epsilon,
\end{equation}
and then perform the integration over $\tau_i \in (-\infty,0]$, which will transform the integrals into twisted integrals over $x_i$'s of the following form:
\begin{equation}
    \hat{\psi}_{n}^{(L)} =  C(\epsilon,Y_e) \int_{\mathbb{R}^n \geq 0} u \cdot \psi^{(L)}_n , 
    \qquad u=\prod_{i=1}^n \, x_i^\epsilon ,
    \quad \psi^{(L)}_n = \tilde{\psi}^{(L)}_n \,  \rmd x_1 \wedge \ldots \rmd x_n.
\end{equation}
Here $C(\epsilon,Y_e) = \left(\frac{\lambda \, \ii^{\epsilon-1}}{\Gamma(1+\epsilon)}\right)^n \prod_{e \in \mathcal{E}} \frac{1}{2Y_e} $ is universal. We will always omit it from now on. And $\psi^{(L)}_n$ is a top form in the $n$-dimensional $x$-space. Instead of performing the time integration every time, $\psi^{(L)}_n$ can easily be written down by considering the sum of all compatible tubings of the diagram. Taking $2$-site chain for example, we have 3 tubings:

\begin{equation}\label{eq:2chaintubings}
\begin{aligned}
    & B_1 = \raisebox{-0.2em}{\begin{tikzpicture}
		\coordinate (X2) at (1,0);
		\coordinate (X3) at (2,0);
		\draw[line width=1.5pt] (X2)--(X3);
		\path[fill=black] (X2) circle[radius=0.1];
		\path[fill=black] (X3) circle[radius=0.1];
    {};
    \draw[line width=0.6pt] (1, 0) circle (2mm);
\end{tikzpicture}} = X_1 + x_1 + Y, \quad B_2 = \raisebox{-0.2em}{\begin{tikzpicture}
		\coordinate (X2) at (1,0);
		\coordinate (X3) at (2,0);
		\draw[line width=1.5pt] (X2)--(X3);
		\path[fill=black] (X2) circle[radius=0.1];
		\path[fill=black] (X3) circle[radius=0.1];
    {};
    \draw[line width=0.6pt] (2, 0) circle (2mm);
\end{tikzpicture}} = X_2 + x_2 + Y,\\
& B_3 = \raisebox{-0.5em}{\begin{tikzpicture}
		\coordinate (X2) at (1,0);
		\coordinate (X3) at (2,0);
		\draw[line width=1.5pt] (X2)--(X3);
		\path[fill=black] (X2) circle[radius=0.1];
		\path[fill=black] (X3) circle[radius=0.1];
        \node [
        draw, color=black, line width=0.6pt,
        rounded rectangle,
        minimum height = 1.3em,
        minimum width = 4.9em,
        rounded rectangle arc length = 180,
    ] at (1.5,0)
    {};
\end{tikzpicture}} = X_1+x_1+X_2+x_2,
\end{aligned}
\end{equation}
and the integrand is written as \footnote{By a slight abuse of notation, we will also denote the integrands and integrals by the corresponding tubing diagram.}
\begin{equation}\label{eq:2chainintegrand}
\psi^{(0)}_{\text{2-site}} =  \raisebox{-0.5em}{\begin{tikzpicture}
		\coordinate (X2) at (1,0);
		\coordinate (X3) at (2,0);
		\draw[line width=1.5pt] (X2)--(X3);
		\path[fill=black] (X2) circle[radius=0.1];
		\path[fill=black] (X3) circle[radius=0.1];
        \node [
        draw, color=black, line width=0.6pt,
        rounded rectangle,
        minimum height = 1.3em,
        minimum width = 4.9em,
        rounded rectangle arc length = 180,
    ] at (1.5,0)
    {};
    \draw[line width=0.6pt,blue] (1, 0) circle (2mm);
    \draw[line width=0.6pt,red] (2, 0) circle (2mm);
\end{tikzpicture}} = \frac{2Y \, \rmd x_1 \wedge \rmd x_2}{{\color{blue}B_1} {\color{red} B_2} B_3} =\frac{2Y \, \rmd x_1 \wedge \rmd x_2}{{\color{blue}(X_1+x_1+Y)}{\color{red}(X_2+x_2+Y)}(X_1+X_2+x_1+x_2)}.
\end{equation}
Note that the integral is defined up to a total derivative
\begin{equation}
    \int u \cdot \psi^{(L)}_n =  \int u \cdot (\psi^{(L)}_n 
    + \nabla_\omega \, \xi),
\end{equation}
where $\nabla_\omega(\bullet)= (\rmd + \omega \, \wedge)(\bullet) = (\rmd + \rmd \log u \, \wedge)(\bullet)$ and $\xi$ is an $(n-1)$-form. Then we are essentially considering the twisted cohomology group of $\mathbb{C}^n \backslash \bigcup_i (x_i=0) \backslash \bigcup_j (B_j=0)$, \textit{i.e.}
\begin{equation}
\left| \psi^{(L)}_n \right. \rangle \in {\left\{\psi|\ \nabla_\omega \, \psi =0\right\}} / {\left\{\nabla_\omega \, \xi \right\}}, \quad \omega = \rmd \log u.
\end{equation}
The differential equation (DE) is a powerful tool for studying this kind of integral~\cite{Arkani-Hamed:2023bsv, Arkani-Hamed:2023kig,baumann2025geometry,he2025differential,chen2025multivariate,He:2024olr}. Remarkably, the differential equation for wavefunction coefficients takes a canonical form
\begin{equation}\label{eq:DE}
  \rmd \, \vec{\psi} = \epsilon \, \mathbf{A}(X_i,Y_i) \, \vec{\psi}, \quad \mathbf{A} = \sum_i \mathbf{c}_i \, \rmd \log \alpha_i,
\end{equation}
where $\vec{\psi}$ is a set of basis of the twisted integral, $\mathbf{A}(X_i,Y_i)$ is a matrix of $\rmd \log$ 1-forms and free of $\epsilon$, $\mathbf{c}_i$ are matrices of rational numbers, and we call $\alpha_i$ letters and the collection $\{\alpha_i\}$ alphabet~\cite{Capuano:2026pgq,Paranjape:2026htn,Mazloumi:2025pmx}. Generally speaking, the solution to the DE is given by generalized hypergeometric functions. In the small-$\epsilon$ regime, one may solve the equation perturbatively order by order in $\epsilon$ (equivalently, by expanding the generalized hypergeometric functions around $\epsilon\to 0$). At each order, the integrated result is expressed in terms of multiple polylogarithms (MPLs). These functions exhibit a rich algebraic structure, known as the coaction, which has been generalized to the case of generalized hypergeometric functions in~\cite{Abreu:2019wzk}, as we review in the next section.

\section{Coaction on Multipolylogarithms and Hypergeometric Functions}\label{sec:iii}
In this subsection, we briefly review the coaction on multipolylogarithms (MPLs) and hypergeometric functions. We only provide the needed information and set up our notation. For more detailed discussion, one can refer to~\cite{Duhr:2012fh} and references therein.

Like many cases of multi-loop Feynman integrals, the twisted integral can be evaluated in terms of MPLs, defined by the iterated integrals
\begin{equation}
    G(a_1,\ldots,a_n;z) = \int_0^z \frac{\rmd t}{(t-a_1)} G(a_2,\ldots,a_n;t),
\end{equation}
where the $a_i$ and $z$ are either constants or functions of certain variables. If all $a_i=0$, the integral diverges, and we define
\begin{equation}
    G(\vec{0}_n;z)= \frac{1}{n!}\log^nz\,,\qquad \vec{0}_n=(\underbrace{0,\ldots,0}_{n\textrm{ times}})\,.
\end{equation}
We call an MPL is of weight-$n$ if it is defined by an $n$-fold iterated integral. The space of all weight-$n$ MPL is $\mathcal{H}_n$, including lower weight MPLs multiplied by transcendental constants, \textit{e.g.} $\pi$ and $\zeta_n$. Roughly speaking, the coaction of an MPL is a map
\begin{equation}
    \mathcal{H}_n \stackrel{\Delta}{\longrightarrow} \bigoplus_{p+q=n}\mathcal{H}_p\otimes (\mathcal{H}_q \text{ mod } \pi),
\end{equation}
which takes an MPL to a linear combination of tensor products of lower weight MPLs. Note that the second entry is defined modulo $\pi$. Two simple examples are as follows,
    \begin{align}
        & \Delta(\log^n z)= \sum_{k=0}^n\binom{n}{k}\log^{n-k} z\otimes \log^k z,\\
        & \Delta(\textrm{Li}_n(z)) = 1\otimes\textrm{Li}_n(z) + \sum_{k=0}^{n-1}\textrm{Li}_{n-k}(z)\otimes\frac{\log^kz}{k!}.\label{eq:coactionofLin}
    \end{align}
For the transcendental constants $\zeta_n$ and $\pi$, we also have the following definition of coaction:
\begin{equation}
\Delta(\zeta_n) = \left\{
\begin{aligned}
    & \zeta_n\otimes 1+ 1\otimes \zeta_n\,,\qquad & n \textrm{ odd}\,, \\
    & \zeta_n\otimes 1\,,\qquad & n \textrm{ even}\,.
\end{aligned}
\right.
\end{equation}
and
\begin{equation}
    \Delta(\ii \, \pi) = (\ii \, \pi) \otimes 1.
\end{equation}
For later convenience, we write
\begin{equation}
    \Delta= \sum_{p+q=n}\Delta_{p,q},
\end{equation}
where $\Delta_{p,q}$ denotes the $\mathcal{H}_p\otimes\mathcal{H}_q$ part in the coaction. And we can further take coaction of a coaction. There are two seemingly different ways to do so, \textit{i.e.} $(\Delta \otimes \text{id})$ and $(\text{id} \otimes \Delta)$, but coassociativity ensures that their results are the same. So we have a unique way ``decompose" a weight-$n$ MPL. We will denote a particular decomposition into weight $\{n_1,n_2,\ldots,n_j\}$ by $\Delta_{n_1,n_2,\ldots,n_j}$. Note that the total differential is the coaction $\Delta_{n,1}$ modulo $\pi$, and the symbol is simply $\Delta_{1,1,\ldots,1}$ modulo $\pi$ with $n$ subscripts is the symbol.

At each order of $\epsilon$ in the perturbative solution of DE~\eqref{eq:DE}, the coaction is calculated as mentioned above. We can also resum the result, and come to the notion of coaction on the resumed function. As one can check order by order: $\Delta(m^{-2\epsilon}) \,= m^{-2\epsilon}\otimes m^{-2\epsilon}$. We then give a more non-trivial example for hypergeometric functions ${}_2F_1$, defined by
\begin{equation}
\label{2F1}
 {}_2F_1(\alpha,\beta;\gamma;x)=\frac{\Gamma(\gamma)}{\Gamma(\alpha)\Gamma(\gamma-\alpha)}
 \int_0^1du\,u^{\alpha-1}(1-u)^{\gamma-\alpha-1}(1-ux)^{-\beta}\, .
\end{equation}
We can expand an explicit hypergeometric function as
\begin{equation}
    {}_2F_1(-\epsilon,1;1-\epsilon;x) = 1-\sum_{n=1}^{\infty}\epsilon^n\,\textrm{Li}_n(x) 
= 1-F(\epsilon,x)\,.
\end{equation}
Considering Eq.~\eqref{eq:coactionofLin}, the resummed coaction can be written as
\begin{equation}
    \Delta\left[F(\epsilon,x)\right] =  1\otimes F(\epsilon,x) + F(\epsilon,x)\otimes x^{\epsilon}\,.
\end{equation}
Then it is easy to see that
\begin{equation}
    \Delta\left[ {}_2F_1(-\epsilon,1;1-\epsilon;x) \right] = 1 \otimes {}_2F_1(-\epsilon,1;1-\epsilon;x) - F(\epsilon,x)\otimes x^\epsilon \, .
\end{equation}
More generally, the coaction on a twisted integral can be written as~\cite{Abreu:2019wzk}
\begin{equation}\label{formulaCoactionTwisted}
 \Delta \left(\int _{\Gamma} u\cdot \psi \right) =\sum _{i,j}\left[\mathbf{C}^{-1}(\vec{\Omega } ,\Omega (\vec{\gamma }) ;u)\right]_{ij} \cdot \int _{\Gamma} u \cdot \Omega _{i} \otimes \int _{\gamma _{j}} u\cdot \psi,
\end{equation}
where the left and right entry are given by master forms $\vec{\Omega}$ (which can be chosen as the canonical forms of the bounded regions) and the dual contours $\vec{\gamma}$ respectively. And $\mathbf{C}(\vec{\Omega } ,\Omega (\vec{\gamma }) ;u)$ is the intersection number matrix, where the intersection number is
\begin{equation}
    \mathbf{C}(\vec{\Omega } ,\Omega (\vec{\gamma }) ;u)_{ij} =  \langle \Omega(\gamma_j) | \Omega_i \rangle,
\end{equation}
and $\Omega(\gamma_j)$ denotes a member of the dual cohomology or the relative twisted cohomology, defined as the direct sum of the twisted cohomology on boundaries.
\begin{equation}
    H^{n}(M,B,\nabla _{-\omega}) = \overset{n}{\underset{p=0}{\bigoplus }}\overset{}{\underset{|J|=p}{\bigoplus }} \ H^{n-|J|}( M\cap B_{J} ,\nabla _{-\omega } |_{J}),
\end{equation}
where $M=\mathbb{C}^n \backslash \bigcup_i (x_i=0)$, and $B_J$ denotes a collection of \textit{untwisted} boundaries with index set $J$. To define an element in the relative twisted cohomology, we introduce the coboundary symbol $\delta_J$ as in \cite{de2024physicalbasiscosmologicalcorrelators}, which means we consider the integral/integrand on the boundary formed by intersection of $B_J$. So an element in the relative twisted cohomology will looks like $ \delta_J (\left. \Omega \right|_J)$, where $\left. \Omega \right|_J$ is an element in the twisted cohomology on boundary $B_J$. 

The intersection number in relative twisted cohomology can be defined as that in ordinary twisted cohomology on boundary $J$, which can be evaluated by \cite{matsumoto1998kforms,De:2023xue}:
\begin{equation}\label{eq:intnum}
    \langle \delta_J (\Omega_J) | \psi \rangle = \langle \Omega_J | \text{Res}_J [\psi] \rangle = \sum_{z_0 \in \text{Int}_J} \frac{\text{Res}_{z_0 = 0}[\Omega_J] \text{Res}_{z_0 = 0}\left[\text{Res}_J [\psi]\right] }{\prod_{i=1}^{n-|J|} \text{Res}_{z_0=0} [\omega]} \, ,
\end{equation}
where $\text{Int}_J$ are the maximal intersection points of the twisted hyperplanes and the plane at infinity.
Since the computation of intersection number only involves $\rmd \log$ forms (on some boundary), the residue can be directly recognized as some lower-dimensional forms, as is shown in the examples in the next section.

\section{Diagrammatic Coaction of 2-site Chain}\label{sec:iv}

Before delving into the details of 2-site chain, we consider a much simpler example, the 1-site diagram. This is the simplest wavefuntion coefficient in FRW cosmology. We can get the integrated result by direct integration:
\begin{equation}
J_{\{1\}}=\raisebox{-0.25em}{\begin{tikzpicture}
\draw[fill] (0, 0) circle (.5mm);
\draw[line width=0.6pt] (0, 0) circle (2mm);
\end{tikzpicture}} = \int_{\mathbb{R}_{\geqslant  0}} \rmd x \, x^\epsilon \, \frac{1}{B_1} = \int_{\mathbb{R}_{\geqslant  0}} \rmd x \, x^\epsilon \, \frac{1}{x+X} = -\pi X^\epsilon \csc(\pi \epsilon)\,.
\end{equation}
The corresponding twisted cohomology group is 1-dimensional, and we can choose the basis as
\begin{equation}
   \Omega_1 = \rmd \log (x+X), \quad \gamma_1 = \delta_{x+X}(1).
\end{equation}
And the intersection number is easy to calculate
\begin{equation}
    C_{1,1}=\langle \gamma_1 | \Omega_1 \rangle = \langle \delta_{x+X}(1) | \rmd \log (x+X) \rangle = \langle 1 | \text{Res}_{x+X=0}[\rmd \log (x+X)] \rangle = 1.
\end{equation}
So we simply have the following expression for its coaction
\begin{equation}
    \Delta \left(\int _{\mathbb{R}_{\geqslant 0}} x^\epsilon \cdot \mathrm{d}\log(x+X) \right) =\underbrace{\int _{\mathbb{R}_{\geqslant 0}} x^\epsilon \cdot \mathrm{d}\log(x+X)}_{J_{\{1\}}} \otimes \underbrace{\int _{\gamma_1} x^\epsilon \cdot \mathrm{d}\log(x+X)}_{\mathcal{C}_{\{1\}} J_{\{1\}}}\, ,
\end{equation}
where the first entry is the same as the original integral, and $\mathcal{C}_{\{1\}}$ in the second integral means we ``cut" the denominator $B_1$, \textit{i.e.} taking residue at $B_1=0$. The second entry evaluates to
\begin{equation}
    \mathcal{C}_{\{1\}} J_{\{1\}} = \left. x^\epsilon \right|_{x+X=0} \text{Res}_{x+X=0} \rmd \log (x+X) = (-X)^\epsilon.
\end{equation}
Note that the second entry in the coaction is defined mod $\pi$, so we can simply eliminate the minus sign, \textit{i.e.} $\mathcal{C}_{\{1\}} J_{\{1\}} \cong X^\epsilon$. Thus we have \footnote{The 1-site result is simple enough to allow a direct computation of coation: $\Delta \left( -\pi X^{\epsilon }\csc( \pi \epsilon )\right) =\Delta ( -\pi ) \cdot \Delta \left( X^{\epsilon }\right) \cdot \Delta (\csc( \pi \epsilon )) =-\pi X^{\epsilon }\csc( \pi \epsilon ) \otimes X^{\epsilon }$, where we have used $\displaystyle \Delta ( \pi ) =\pi \otimes 1,\ \Delta \left( X^{\epsilon }\right) =X^{\epsilon } \otimes X^{\epsilon } ,\ \Delta (\csc( \pi \epsilon )) =\csc( \pi \epsilon ) \otimes 1$, and comultiplication $\displaystyle \Delta ( a\cdot b) =\Delta ( a) \cdot \Delta ( b)$. }
\begin{equation}
    \Delta \left( -\pi X^{\epsilon }\csc( \pi \epsilon )\right) =-\pi X^{\epsilon }\csc( \pi \epsilon ) \otimes X^{\epsilon},
\end{equation}
or diagrammatically
\begin{equation}
    \Delta \,
    \raisebox{-0.1em}{\begin{tikzpicture}
    \draw[fill] (0, 0) circle (.5mm);
    \draw[line width=0.6pt] (0, 0) circle (2mm);
    \end{tikzpicture}} = \raisebox{-0.1em}{\begin{tikzpicture}
    \draw[fill] (0, 0) circle (.5mm);
    \draw[line width=0.6pt] (0, 0) circle (2mm);
    \end{tikzpicture}} \otimes 
    \raisebox{-0.1em}{\begin{tikzpicture}
    \draw[fill] (0, 0) circle (.5mm);
    \draw[line width=0.6pt,dashed] (0, 0) circle (2mm);
    \end{tikzpicture}} \,\, ,
\end{equation}
where the dashed tubing is understood to integrate on the corresponding cut/boundary. Then we will next see how this structure generalize to 2-site chain.

\subsection{The Explicit Calculation}
The tubings/hyperplanes and the integrand for 2-site chain are already defined in \eqref{eq:2chaintubings} and \eqref{eq:2chainintegrand}. And we put the hyperplanes here again for convenience, and two twisted hyperplanes $T_1$ and $T_2$ are included:

\begin{equation}\label{eq:2site T B hyperplanes}
\begin{aligned}
    & T_1 = x_1, \quad T_2 = x_2, \\
    & B_1 = \raisebox{-0.2em}{\begin{tikzpicture}
		\coordinate (X2) at (1,0);
		\coordinate (X3) at (2,0);
		\draw[line width=1.5pt] (X2)--(X3);
		\path[fill=black] (X2) circle[radius=0.1];
		\path[fill=black] (X3) circle[radius=0.1];
    {};
    \draw[line width=0.6pt] (1, 0) circle (2mm);
\end{tikzpicture}} = X_1 + x_1 + Y, \quad B_2 = \raisebox{-0.2em}{\begin{tikzpicture}
		\coordinate (X2) at (1,0);
		\coordinate (X3) at (2,0);
		\draw[line width=1.5pt] (X2)--(X3);
		\path[fill=black] (X2) circle[radius=0.1];
		\path[fill=black] (X3) circle[radius=0.1];
    {};
    \draw[line width=0.6pt] (2, 0) circle (2mm);
\end{tikzpicture}} = X_2 + x_2 + Y,\\
& B_3 = \raisebox{-0.5em}{\begin{tikzpicture}
		\coordinate (X2) at (1,0);
		\coordinate (X3) at (2,0);
		\draw[line width=1.5pt] (X2)--(X3);
		\path[fill=black] (X2) circle[radius=0.1];
		\path[fill=black] (X3) circle[radius=0.1];
        \node [
        draw, color=black, line width=0.6pt,
        rounded rectangle,
        minimum height = 1.3em,
        minimum width = 4.9em,
        rounded rectangle arc length = 180,
    ] at (1.5,0)
    {};
\end{tikzpicture}} = X_1+x_1+X_2+x_2,
\end{aligned}
\end{equation}
For later convenience, we choose the following orthogonal pair of $(\vec{\Omega} ,\Omega(\vec{\gamma}))$:
\begin{equation}
    \vec{\Omega} =\begin{pmatrix}
\rmd \log B_1 \wedge \rmd \log B_2 \\
\rmd \log B_1 \wedge \rmd \log B_3\\
\rmd \log B_2 \wedge \rmd \log B_3\\
\rmd  \log B_3 \wedge \rmd \log T_1
\end{pmatrix} ,\ \Omega(\vec{\gamma})=\begin{pmatrix}
\delta _{12}(1)\\
\delta _{13}(1)\\
\delta _{23}(1)\\
\delta _{3} \left(\rmd \log \frac{\left. T_1 \right|_3 }{\left. T_2 \right|_3}\right)
\end{pmatrix}\, .
\end{equation}
Here $\left.T\right|_{3}$ denotes the restriction of the expression $T$ to the hyperplane $B_3=0$, while for a differential form $\omega$, $\left.\omega\right|_{3}$ denotes its pullback to that hyperplane.
For the fourth basis:
\begin{equation}
     \begin{aligned}
& C_{44} = \left\langle \Omega \left( \gamma_4 \right) ,\Omega _{4} \right\rangle =  \left\langle \delta _{3} \left(\rmd \log \frac{\left. T_1 \right|_3 }{\left. T_2 \right|_3}\right), \mathrm{d}\log B_{3} \wedge \rmd \log T_1 \right\rangle \\
& = \left\langle \rmd \log \frac{\left. T_1 \right|_3 }{\left. T_2 \right|_3}, \text{Res}_{B_3=0} ( \mathrm{d}\log B_{3} \wedge \rmd \log T_1 ) \right\rangle
=  \left\langle \rmd \log \frac{\left. T_1 \right|_3 }{\left. T_2 \right|_3}, \rmd \log \left. T_1 \right|_3 \right\rangle =  \frac{1}{\epsilon} \, .
\end{aligned}
\end{equation}
where we have used Eq.~\eqref{eq:intnum}. For other elements in the intersection matrix, it is easy to see that $C_{i,i}=1$ and $C_{i,j}=0$ for $i\neq j$. Then, the coaction on 2-site chain is simply written as:
\begin{equation}\label{eq:2sitechaincoaction}
    \Delta \left(\int _{\mathbb{R}_{\geqslant 0}^{2}} u\cdot \psi^{(0)}_{\text{2-site}} \right) =\sum _{i=1,\cdots ,4} C_{i,i}^{-1} \int _{\mathbb{R}_{\geqslant 0}^{2}} u\cdot \Omega _{i} \otimes \int _{\gamma _{i}} u\cdot \psi^{(0)}_{\text{2-site}}.
\end{equation}
or rewrite it in another notation:
\begin{equation}\label{eq:2siteco}
\Delta [J_{\{1,2,3\}} ]=\sum _{X=\{1,2\},\{1,3\},\{2,3\},\{3\}} \epsilon ^{2-|X|} \cdot J_{X} \otimes \mathcal{C}_{X} J_{\{1,2,3\}}\, ,
\end{equation}
where we denote the integral with poles $\{B_i\}_{i \in X}$ by $J_X$ and $\mathcal{C}_Y$ meas we ``cut" all the poles in $\{B_j\}_{j \in Y}$. Eq.~\eqref{eq:2sitechaincoaction} and Eq.~\eqref{eq:2siteco} is expected to be consistent with the integrated result for 2-site chain given in \cite{Arkani-Hamed:2023kig}. To check this, we will calculate the integrated result of the coaction, and expand each of them in small $\epsilon$. Also, we can expand the integrated result for 2-site chain, then the normal MPL coaction can be applied order by order in $\epsilon$, after which, we will find consistency. 

We first compute the result of left entries in the coaction. For $ \Omega _{1} =\mathrm{d}\log B_1 \wedge \rmd \log B_2 $, we have:
\begin{equation}
\begin{aligned}
J_{\{1,2\}}  =\int _{\mathbb{R}_{\geqslant 0}^{2}}( x_{1} x_{2})^{\epsilon } \cdot \frac{\mathrm{d} x_{1}\mathrm{d} x_{2}}{( x_{1} +X_{1} +Y)( x_{1} +X_{1} +Y)} =\pi ^{2}\csc^{2} (\pi \epsilon )( X_{1} +Y)^{\epsilon }( X_{2} +Y)^{\epsilon }.
\end{aligned}
\end{equation}
Similarly we can work out the left entries of the components $\Omega_2,\Omega_3$:
\begin{equation}
\begin{aligned}
J_{\{1,3\}} & =\frac{1}{4}\sqrt{\pi }\csc (\pi \epsilon )\Bigl[\frac{4^{-\epsilon }( X_{1} +X_{2})^{2\epsilon +1} \Gamma \left( -\epsilon -\frac{1}{2}\right) \Gamma (\epsilon +1)\ _{2} F_{1}\left( 1,\epsilon +1;2(\epsilon +1);\frac{X_{1} +X_{2}}{X_{2} -Y}\right)}{X_{2} -Y}\\
 & -2\pi ^{3/2}\csc (\pi \epsilon )( X_{2} -Y)^{\epsilon }\left(\sec (\pi \epsilon )\left( -\frac{X_{1} +Y}{X_{2} -Y}\right)^{\epsilon }( X_{2} -Y)^{\epsilon } -2( X_{1} +Y)^{\epsilon }\right)\Bigr],
\end{aligned}
\end{equation}
\begin{equation}
    \begin{aligned}
J_{\{2,3\}} & =-\pi \csc (\pi \epsilon )\Bigl[\frac{2^{-2(\epsilon +1)}( X_{1} +X_{2})^{2\epsilon +1} \Gamma \left( -\epsilon -\frac{1}{2}\right) \Gamma (\epsilon +1)\ _{2} F_{1}\left( 1,\epsilon +1;2(\epsilon +1);\frac{X_{1} +X_{2}}{Y+X_{2}}\right)}{\sqrt{\pi }( X_{2} +Y)}\\
 & -\pi \csc (2\pi \epsilon )\left(\frac{Y-X_{1}}{X_{2} +Y}\right)^{\epsilon }( X_{2} +Y)^{2\epsilon }\Bigr].
\end{aligned}
\end{equation}
Note that, for $\Omega_4$, we have only one $B$-type pole, but with one extra $x_1$ pole:
\begin{equation}
\begin{aligned}
J_{\{3\}} =\int _{\mathbb{R}_{\geqslant 0}}( x_{1} x_{2})^{\epsilon } \cdot \frac{ \mathrm{d} x_{1}\mathrm{d} x_{2}}{x_1 ( x_{1} +x_{2} +X_{1} +X_{2})}
 = \frac{\pi (X_{1} +X_{2} )^{2\epsilon }\csc (\pi \epsilon )\Gamma (-2\epsilon )\Gamma (\epsilon +1)}{\Gamma (1-\epsilon )}.
\end{aligned}
\end{equation}

Then we compute the right entries. For $\Omega(\gamma_1)=\delta _{12}(1)$, we have:
\begin{equation}
\begin{aligned}
\mathcal{C}_{\{1,2\}} J_{\{1,2,3\}} & = \mathcal{C}_{\{1,2\}}  \int  u\cdot \psi^{(0)}_{\text{2-site}} =u|_{12} \cdot \mathrm{Res}_{12}(\rmd \log B_1 \wedge \rmd \log B_2) \\
 & = x_{1}^{\epsilon } |_{B_{1} =0} \cdot x_{2}^{\epsilon } |_{B_{2} =0} = ( -X_{1} -Y)^{\epsilon } \cdot ( -X_{2} -Y)^{\epsilon}\, .
\end{aligned}
\end{equation}
Note that right entries are defined modulo $ \pi $, so we have:
\begin{equation}
    \mathcal{C}_{\{1,2\}} J_{\{1,2,3\}} =( X_{1} +Y)^{\epsilon }( X_{2} +Y)^{\epsilon }.
\end{equation}
Likewise, we can obtain the right entries for $\Omega(\gamma_2)$ and $\Omega(\gamma_3)$:
\begin{equation}
    \mathcal{C}_{\{1,3\}} J_{\{1,2,3\}} =( X_{1} +Y)^{\epsilon }( X_{2} -Y)^{\epsilon },
\end{equation}
\begin{equation}
    \mathcal{C}_{\{2,3\}} J_{\{1,2,3\}} =( X_{1} -Y)^{\epsilon }( X_{2} +Y)^{\epsilon }.
\end{equation}
For the fourth component, it is a bit complicated since it is a next-to-maximal cut:
\begin{equation}
    \begin{aligned}
\mathcal{C}_{\{3\}} J_{\{1,2,3\}} & =\int u|_{3} \cdot \mathrm{Res}_{3}\psi^{(0)}_{\text{2-site}}\\
 & =\int _{0}^{-X_{1} -X_{2}} x^{\epsilon }( -( x +X_{1} +X_{2}))^{\epsilon } \cdot \left(\frac{\mathrm{d} x}{x +X_{1} +Y} +\frac{\mathrm{d} x}{x +X_{2} +Y}\right)\\
 & =( -X_{1} -X_{2})^{2\epsilon +1}\left( -\Gamma (\epsilon +1)^{2}\right)\Bigl[\frac{\ _{2}\tilde{F}_{1}\left( 1,\epsilon +1;2(\epsilon +1);\frac{X_{1} +X_{2}}{X_{2} -Y}\right)}{Y-X_{2}}\\
 & +\frac{\ _{2}\tilde{F}_{1}\left( 1,\epsilon +1;2(\epsilon +1);\frac{X_{1} +X_{2}}{Y+X_{2}}\right)}{X_{2} +Y}\Bigr]\, ,
\end{aligned}
\end{equation}
where the regularized hypergeometric function $\displaystyle _{2}\tilde{F}_{1}( a,b;c;x) = \ _{2} F_{1}( a,b;c;x) /\Gamma ( c)$.

We have verified up to $O(\epsilon^1)$ that of the above coaction formula is consistent with the integrated result. And we also check the \textit{coassociativity} of the coaction. See Appendix \ref{app:A} for more details.

\subsection{The Diagrammatic Interpretation}
Taking a closer look on each term in Eq.~\eqref{eq:2sitechaincoaction} and Eq.~\eqref{eq:2siteco}, the left entries turn out to be certain ``subtopology" of the original integral, \textit{i.e.} only a subset of untwisted hyperplanes appear in the denominator. And for the right entries, it is the integral evaluated on the ``cut" defined by the same subset of hyperplanes for the left entry. The coaction then allows for a natural diagrammatic interpretation:
\begin{equation}
\begin{aligned}
    \Delta  \,\,\, \raisebox{-0.3em}{\begin{tikzpicture}
		\coordinate (X2) at (1,0);
		\coordinate (X3) at (2,0);
		\draw[line width=1.5pt] (X2)--(X3);
		\path[fill=black] (X2) circle[radius=0.1];
		\path[fill=black] (X3) circle[radius=0.1];
        \node [
        draw, color=black, line width=0.6pt,
        rounded rectangle,
        minimum height = 1.3em,
        minimum width = 4.9em,
        rounded rectangle arc length = 180,
    ] at (1.5,0)
    {};
    \draw[line width=0.6pt,black] (1, 0) circle (2mm);
    \draw[line width=0.6pt,black] (2, 0) circle (2mm);
\end{tikzpicture}}  & = \raisebox{-0.3em}{\begin{tikzpicture}
		\coordinate (X2) at (1,0);
		\coordinate (X3) at (2,0);
		\draw[line width=1.5pt] (X2)--(X3);
		\path[fill=black] (X2) circle[radius=0.1];
		\path[fill=black] (X3) circle[radius=0.1];
        \node [
        draw, color=white, line width=0.6pt,
        rounded rectangle,
        minimum height = 1.3em,
        minimum width = 4.9em,
        rounded rectangle arc length = 180,
    ] at (1.5,0)
    {};
    \draw[line width=0.6pt,black] (1, 0) circle (2mm);
    \draw[line width=0.6pt,black] (2, 0) circle (2mm);
\end{tikzpicture}}  \otimes \,\, \raisebox{-0.3em}{\begin{tikzpicture}
		\coordinate (X2) at (1,0);
		\coordinate (X3) at (2,0);
		\draw[line width=1.5pt] (X2)--(X3);
		\path[fill=black] (X2) circle[radius=0.1];
		\path[fill=black] (X3) circle[radius=0.1];
        \node [
        draw, color=black, line width=0.6pt,
        rounded rectangle,
        minimum height = 1.3em,
        minimum width = 4.9em,
        rounded rectangle arc length = 180,
    ] at (1.5,0)
    {};
    \draw[line width=0.6pt,black,dashed] (1, 0) circle (2mm);
    \draw[line width=0.6pt,black,dashed] (2, 0) circle (2mm);
\end{tikzpicture}}\\
& + \raisebox{-0.3em}{\begin{tikzpicture}
		\coordinate (X2) at (1,0);
		\coordinate (X3) at (2,0);
		\draw[line width=1.5pt] (X2)--(X3);
		\path[fill=black] (X2) circle[radius=0.1];
		\path[fill=black] (X3) circle[radius=0.1];
        \node [
        draw, color=black, line width=0.6pt,
        rounded rectangle,
        minimum height = 1.3em,
        minimum width = 4.9em,
        rounded rectangle arc length = 180,
    ] at (1.5,0)
    {};
    \draw[line width=0.6pt,black] (1, 0) circle (2mm);
\end{tikzpicture}}  \otimes \,\, \raisebox{-0.3em}{\begin{tikzpicture}
		\coordinate (X2) at (1,0);
		\coordinate (X3) at (2,0);
		\draw[line width=1.5pt] (X2)--(X3);
		\path[fill=black] (X2) circle[radius=0.1];
		\path[fill=black] (X3) circle[radius=0.1];
        \node [
        draw, color=black, line width=0.6pt, dashed,
        rounded rectangle,
        minimum height = 1.3em,
        minimum width = 4.9em,
        rounded rectangle arc length = 180,
    ] at (1.5,0)
    {};
    \draw[line width=0.6pt,black,dashed] (1, 0) circle (2mm);
    \draw[line width=0.6pt,black] (2, 0) circle (2mm);
\end{tikzpicture}}\\
& + \raisebox{-0.3em}{\begin{tikzpicture}
		\coordinate (X2) at (1,0);
		\coordinate (X3) at (2,0);
		\draw[line width=1.5pt] (X2)--(X3);
		\path[fill=black] (X2) circle[radius=0.1];
		\path[fill=black] (X3) circle[radius=0.1];
        \node [
        draw, color=black, line width=0.6pt,
        rounded rectangle,
        minimum height = 1.3em,
        minimum width = 4.9em,
        rounded rectangle arc length = 180,
    ] at (1.5,0)
    {};
    \draw[line width=0.6pt,black] (2, 0) circle (2mm);
\end{tikzpicture}}  \otimes \,\, \raisebox{-0.3em}{\begin{tikzpicture}
		\coordinate (X2) at (1,0);
		\coordinate (X3) at (2,0);
		\draw[line width=1.5pt] (X2)--(X3);
		\path[fill=black] (X2) circle[radius=0.1];
		\path[fill=black] (X3) circle[radius=0.1];
        \node [
        draw, color=black, line width=0.6pt,dashed,
        rounded rectangle,
        minimum height = 1.3em,
        minimum width = 4.9em,
        rounded rectangle arc length = 180,
    ] at (1.5,0)
    {};
    \draw[line width=0.6pt,black] (1, 0) circle (2mm);
    \draw[line width=0.6pt,black,dashed] (2, 0) circle (2mm);
\end{tikzpicture}}\\
& + \epsilon \, \, \raisebox{-0.3em}{\begin{tikzpicture}
		\coordinate (X2) at (1,0);
		\coordinate (X3) at (1.2,0);
		\draw[line width=1.5pt] (X2)--(X3);
		\path[fill=black] (X2) circle[radius=0.1];
		\path[fill=black] (X3) circle[radius=0.1];
        \node [
        draw, color=black, line width=0.6pt,
        rounded rectangle,
        minimum height = 1.3em,
        minimum width = 2em,
        rounded rectangle arc length = 180,
    ] at (1.1,0)
    {};
\end{tikzpicture}}  \otimes \,\, \raisebox{-0.3em}{\begin{tikzpicture}
		\coordinate (X2) at (1,0);
		\coordinate (X3) at (2,0);
		\draw[line width=1.5pt] (X2)--(X3);
		\path[fill=black] (X2) circle[radius=0.1];
		\path[fill=black] (X3) circle[radius=0.1];
        \node [
        draw, color=black, line width=0.6pt,dashed,
        rounded rectangle,
        minimum height = 1.3em,
        minimum width = 4.9em,
        rounded rectangle arc length = 180,
    ] at (1.5,0)
    {};
    \draw[line width=0.6pt,black] (1, 0) circle (2mm);
    \draw[line width=0.6pt,black] (2, 0) circle (2mm);
\end{tikzpicture}}
\end{aligned}
\end{equation}
Furthermore, we can interpret the left entries as some specific time integrals~\cite{he2025differential}, defined in Eq.~\eqref{eq:timedecomp}. Explicitly, using Eq.~\eqref{eq:timepowerrep} to calculate each piece of a specific time-order, we will find:
\begin{equation}
    \begin{aligned}
        \raisebox{-0.5em}{\begin{tikzpicture}
		\coordinate (X2) at (1,0);
		\coordinate (X3) at (2,0);
		\draw[line width=1.5pt] (X2)--(X3);
		\path[fill=black] (X2) circle[radius=0.1];
		\path[fill=black] (X3) circle[radius=0.1];
        \node [
        draw, color=white, line width=0.6pt,
        rounded rectangle,
        minimum height = 1.3em,
        minimum width = 4.9em,
        rounded rectangle arc length = 180,
    ] at (1.5,0)
    {};
    \draw[line width=0.6pt,black] (1, 0) circle (2mm);
    \draw[line width=0.6pt,black] (2, 0) circle (2mm);
\end{tikzpicture}} = \,\, \raisebox{0em}{\begin{tikzpicture}
	\coordinate (X2) at (1,0);
	\coordinate (X3) at (2,0);
	\draw[line width=1.5pt,dashed] (X2)--(X3);
	\path[fill=black] (X2) circle[radius=0.1];
	\path[fill=black] (X3) circle[radius=0.1];
\end{tikzpicture}} \, , \quad \raisebox{-0.5em}{\begin{tikzpicture}
		\coordinate (X2) at (1,0);
		\coordinate (X3) at (2,0);
		\draw[line width=1.5pt] (X2)--(X3);
		\path[fill=black] (X2) circle[radius=0.1];
		\path[fill=black] (X3) circle[radius=0.1];
        \node [
        draw, color=black, line width=0.6pt,
        rounded rectangle,
        minimum height = 1.3em,
        minimum width = 4.9em,
        rounded rectangle arc length = 180,
    ] at (1.5,0)
    {};
    \draw[line width=0.6pt,black] (1, 0) circle (2mm);
\end{tikzpicture}} = \,\, \raisebox{0em}{\begin{tikzpicture}
				\coordinate (X2) at (1,0);
				\coordinate (X3) at (2,0);
				\draw[line width=1.5pt,->] (X2)--(1.6,0);
				\draw[line width=1.5pt] (1.5,0)--(X3);
				\path[fill=black] (X2) circle[radius=0.1];
				\path[fill=black] (X3) circle[radius=0.1];
		\end{tikzpicture}} \, , \quad \raisebox{-0.5em}{\begin{tikzpicture}
		\coordinate (X2) at (1,0);
		\coordinate (X3) at (2,0);
		\draw[line width=1.5pt] (X2)--(X3);
		\path[fill=black] (X2) circle[radius=0.1];
		\path[fill=black] (X3) circle[radius=0.1];
        \node [
        draw, color=black, line width=0.6pt,
        rounded rectangle,
        minimum height = 1.3em,
        minimum width = 4.9em,
        rounded rectangle arc length = 180,
    ] at (1.5,0)
    {};
    \draw[line width=0.6pt,black] (2, 0) circle (2mm);
\end{tikzpicture}} = 
		\raisebox{0em}{\begin{tikzpicture}
		\coordinate (X2) at (1,0);
		\coordinate (X3) at (2,0);
		\draw[line width=1.5pt,->] (X3)--(1.4,0);
		\draw[line width=1.5pt] (1.5,0)--(X2);
		\path[fill=black] (X2) circle[radius=0.1];
		\path[fill=black] (X3) circle[radius=0.1];
	\end{tikzpicture}} \, .
    \end{aligned}
\end{equation}
And for the fourth term, the left entry can be understood as a ``pinch" of the two sites, which can also be generated from the differential of the time integral~\cite{he2025differential}. Using integration-by-part, we have
\begin{equation}
    \begin{aligned}
\epsilon \, \, \raisebox{-0.3em}{\begin{tikzpicture}
		\coordinate (X2) at (1,0);
		\coordinate (X3) at (1.2,0);
		\draw[line width=1.5pt] (X2)--(X3);
		\path[fill=black] (X2) circle[radius=0.1];
		\path[fill=black] (X3) circle[radius=0.1];
        \node [
        draw, color=black, line width=0.6pt,
        rounded rectangle,
        minimum height = 1.3em,
        minimum width = 2em,
        rounded rectangle arc length = 180,
    ] at (1.1,0)
    {};
\end{tikzpicture}} & \sim \epsilon \int _{\mathbb{R}_{\geqslant 0}^{2}}( x_{1} x_{2})^{\epsilon }\frac{\mathrm{d} x_{1}\mathrm{d} x_{2}}{x_{1}( x_{1} +x_{2} +X_{1} +X_{2})}\\
 & =\epsilon \int _{\mathbb{R}_{\geqslant 0}^{2}}( x_{1})^{\epsilon -1} x_{2}^{\epsilon }\frac{\mathrm{d} x_{1}\mathrm{d} x_{2}}{x_{1} +x_{2} +X_{1} +X_{2}}\\
 & =\epsilon \int _{\mathbb{R}_{\geqslant 0}^{2}}\frac{1}{\epsilon }\left[\frac{\partial }{\partial x_{1}} x_{1}^{\epsilon }\right] x_{2}^{\epsilon }\frac{\mathrm{d} x_{1}\mathrm{d} x_{2}}{x_{1} +x_{2} +X_{1} +X_{2}} \\
 & =-\int _{\mathbb{R}_{\geqslant 0}^{2}} x_{1}^{\epsilon } \cdot \frac{\partial }{\partial x_{1}}\left[ x_{2}^{\epsilon }\frac{1}{x_{1} +x_{2} +X_{1} +X_{2}}\right]\mathrm{d} x_{1}\mathrm{d} x_{2} \\
 & =\int _{\mathbb{R}_{\geqslant 0}^{2}}( x_{1} x_{2})^{\epsilon } \cdot \frac{\mathrm{d} x_{1}\mathrm{d} x_{2}}{( x_{1} +x_{2} +X_{1} +X_{2})^{2}}.
\end{aligned}
\end{equation}
And from the time integral perspective, the ``pinched" site is defined as
\begin{equation}
\begin{aligned}
    \raisebox{-1.5em}{\begin{tikzpicture}
		\coordinate (X2) at (1,0);
		\coordinate (X3) at (1.2,0);
		\draw[line width=1.5pt] (X2)--(X3);
		\path[fill=black] (X2) circle[radius=0.1];
		\path[fill=black] (X3) circle[radius=0.1];
        \node [
        draw, color=white, line width=0.6pt,
        rounded rectangle,
        minimum height = 1.3em,
        minimum width = 2em,
        rounded rectangle arc length = 180,
    ] at (1.1,0)
    {};
    \node[above] at (1.1,0.1) {\small{$\epsilon+\epsilon$}};
    \node[below] at (1.1,-0.1) {\small{$X_1+X_2$}};
\end{tikzpicture}} & =  \int_{-\infty}^{0} \rmd \tau \, \ee^{\ii (X_1+X_2) \tau} (-\tau)^{-1-2\epsilon} = \int_{-\infty}^{0} \rmd \tau \,  \ee^{\ii (X_1+X_2) \tau} (-\tau)^{-1-\epsilon} (-\tau)^{-1-\epsilon} (-\tau)\\
& \sim \int_{-\infty}^{0} \rmd \tau \,  \ee^{\ii (X_1+X_2) \tau} \left( \int_0^\infty \rmd x_1 \, \ee^{\ii x_1 \tau} x_1^\epsilon \right) \left( \int_0^\infty \rmd x_2 \, \ee^{\ii x_2 \tau} x_2^\epsilon \right)  (-\tau)\\
& =\int _{\mathbb{R}_{\geqslant 0}^{2}}( x_{1} x_{2})^{\epsilon } \cdot \frac{\mathrm{d} x_{1}\mathrm{d} x_{2}}{( x_{1} +x_{2} +X_{1} +X_{2})^{2}}.
\end{aligned}
\end{equation}
Thus we have
\begin{equation}
    \epsilon \, \, \raisebox{-0.3em}{\begin{tikzpicture}
		\coordinate (X2) at (1,0);
		\coordinate (X3) at (1.2,0);
		\draw[line width=1.5pt] (X2)--(X3);
		\path[fill=black] (X2) circle[radius=0.1];
		\path[fill=black] (X3) circle[radius=0.1];
        \node [
        draw, color=black, line width=0.6pt,
        rounded rectangle,
        minimum height = 1.3em,
        minimum width = 2em,
        rounded rectangle arc length = 180,
    ] at (1.1,0)
    {};
\end{tikzpicture}} = \raisebox{-1.5em}{\begin{tikzpicture}
		\coordinate (X2) at (1,0);
		\coordinate (X3) at (1.2,0);
		\draw[line width=1.5pt] (X2)--(X3);
		\path[fill=black] (X2) circle[radius=0.1];
		\path[fill=black] (X3) circle[radius=0.1];
        \node [
        draw, color=white, line width=0.6pt,
        rounded rectangle,
        minimum height = 1.3em,
        minimum width = 2em,
        rounded rectangle arc length = 180,
    ] at (1.1,0)
    {};
    \node[above] at (1.1,0.1) {\small{$\epsilon+\epsilon$}};
    \node[below] at (1.1,-0.1) {\small{$X_1+X_2$}};
\end{tikzpicture}}.
\end{equation}

\section{Outlook}\label{sec:v}

In this work, using a two-site example, we study the coaction of cosmological wavefunction coefficients of conformally coupled scalars in FRW background, which turns out to have a diagrammatic interpretation. We also clarified the physical interpretation of the various terms appearing in the coaction, identifying them with time ordered integrals and cuts.

More broadly, the coaction encodes structured information associated with the integrated function. In this sense, it complements both the direct series expansions studied in \cite{fan2025anatomy,Fan:2020xgh} and the differential-equation approach developed in \cite{he2025differential}. In particular, it provides a systematic and general framework for studying the singularity of FRW integrals, generalizing the notion of symbols.

In this work, we only present a two-site example, then a natural question to ask is whether the diagrammatic coaction can be constructed for arbitrary trees and loop integrands of conformally coupled scalars, and whether a similar structure can also be developed for massive scalars. For the massless cases, we leave it to a future work \cite{Fu_Liu_toappear}. And for the massive cases, one may replace the Hankel functions of the second kind appearing in massive propagators by their integral representations, thereby rewriting the problem in terms of a higher-dimensional twisted integral. For example, for the single massive exchange diagram in dS, in close analogy with the twisted-integral representation in \eqref{eq:2chaintubings} and \eqref{eq:2chainintegrand}, one obtains
\begin{equation}
\begin{array}{l}
B_{1}=X_{1}+x_{1}+Ys_{1}, \quad B_{2}=X_{2}+x_{2}+Ys_{2},\\
B_{3}=X_{1}+x_{1}+X_{2}+x_{2}+Y(s_{1}-s_{2}), \quad
B_{4}=X_{1}+x_{1}+X_{2}+x_{2}+Y(s_{2}-s_{1}),
\end{array}
\end{equation}
and the integrand can be written as
\begin{equation}
\psi^{(\nu)}=\psi_{+}^{(\nu)}+\psi_{-}^{(\nu)}+\psi_{0}^{(\nu)},
\end{equation}
where the three terms correspond respectively to the time-ordered, anti-time-ordered, and boundary contributions:
\begin{equation}
\begin{aligned}
\psi_{+}^{(\nu)}
&=-\frac{\mathrm{d}x_{1}\land \mathrm{d}x_{2}\land \mathrm{d}s_{1}\land \mathrm{d}s_{2}}{2YB_{1}B_{3}},\\
\psi_{-}^{(\nu)}
&=-\frac{\mathrm{d}x_{1}\land \mathrm{d}x_{2}\land \mathrm{d}s_{1}\land \mathrm{d}s_{2}}{2YB_{2}B_{4}},\\
\psi_{0}^{(\nu)}
&=e^{2\pi i\nu}\frac{\mathrm{d}x_{1}\land \mathrm{d}x_{2}\land \mathrm{d}s_{1}\land \mathrm{d}s_{2}}{2YB_{1}B_{2}}.
\end{aligned}
\end{equation}
Here,
\begin{equation}
u=(x_{1}x_{2})^{-\epsilon}\bigl(s_{1}s_{2}(s_{1}+2)(s_{2}+2)\bigr)^{\epsilon},
\end{equation}
and the integral is defined by
\begin{equation}
\int_{\mathbb{R}_{\geqslant 0}^{4}} u\cdot \psi^{(\nu)}.
\end{equation}

However, the increase in the dimension of the integral for also introduces a much larger amount of redundancy in the basis. Naively, the hyperplanes in this four-dimensional integration space intersect to produce five boundaries $B_J$, with $J=\{1,2\},\{1,3\},\{2,4\},\{3\},\{4\}$. This would lead to a $20$-dimensional cohomology basis, with four basis elements supported on each boundary. By contrast, the fact that the integral satisfies second-order partial differential equations strongly suggests that the true physical basis should be much smaller.

The main challenge, therefore, is to identify a systematic reduction from this redundant cohomology basis to the true physical one. This is closely related to the mechanism discussed in \cite{de2024physicalbasiscosmologicalcorrelators}, but the present setting is more subtle: in addition to redundant cohomology elements, one also encounters unphysical integrals whose symmetries generate further equivalence relations among the integrands. Understanding how to quotient out these redundancies will be an important step toward extending the coaction framework to massive fields.

\begin{CJK*}{UTF8}{}
\CJKfamily{gbsn}
\acknowledgments
The authors would like to thank Daniel Baumann, Song He (何颂), Hayden Lee, Xuhang Jiang (姜旭航), Austin Joyce, Andrzej Pokraka and Kamran Salehi Vaziri for inspiring and helpful discussions. This work was supported by NSFC Grant No.12225510, 12447101, and by the New Cornerstone Science Foundation.
\end{CJK*}

\appendix

\section{Verification of coaction formula}\label{app:A}
The explicit result for 2-site FRW correlator is:
\begin{equation}\label{eq:2site}
\begin{aligned}
\hat{\psi} _{\text{2-site}}( X_{1} ,X_{2} ,Y) & =\sqrt{\pi } 4^{-\epsilon }( X_{1} +X_{2})^{2\epsilon }\csc (\pi \epsilon )\Gamma \left(\frac{1}{2} -\epsilon \right) \Gamma (\epsilon )\Bigl[ -\ _{2} F_{1}\left( 1,\epsilon ;1-\epsilon ;\frac{Y-X_{2}}{Y+X_{1}}\right)\\
 & -\ _{2} F_{1}\left( 1,\epsilon ;1-\epsilon ;\frac{Y-X_{1}}{Y+X_{2}}\right) +1\Bigr] +\pi ^{2}\csc^{2} (\pi \epsilon )( X_{1} +Y)^{\epsilon }( X_{2} +Y)^{\epsilon }
\end{aligned}
\end{equation}
We now check that our coaction is consistent with the above result. For simplicity, we only present the result of the symbol, which is the $\Delta _{\underbrace{1,\cdots ,1}_{n}}$ component of the coaction for the transcendental weight $n$ function. Expanding around $\epsilon \rightarrow 0$, the symbol up to $\mathcal{O} (\epsilon ^{1} )$ reads
\begin{equation}
\mathcal{S} (\hat{\psi} _{\text{2-site}} )\equiv \mathcal{S}\left( \hat{\psi} _{\text{2-site}}^{(\epsilon^0)}\right) +\epsilon \cdot \mathcal{S}\left( \hat{\psi} _{\text{2-site}}^{(\epsilon^1)}\right) +\mathcal{O} (\epsilon ^{2} ),
\end{equation}
with the $O(\epsilon ^{0} )$ and $O(\epsilon ^{1} )$ component:
\begin{equation}
\mathcal{S}\left( \hat{\psi} _{\text{2-site}}^{(\epsilon^0)}\right) =L_{1} \otimes L_{2} -L_{1} \otimes L_{4} +L_{2} \otimes L_{1} -L_{2} \otimes L_{3} -L_{5} \otimes L_{1} -L_{5} \otimes L_{2} +L_{5} \otimes L_{3} +L_{5} \otimes L_{4},
\end{equation}
\begin{equation}\label{eq:2sitew3}
\begin{aligned}
\mathcal{S}\left( \hat{\psi} _{\text{2-site}}^{(\epsilon^1)}\right) & =L_{1} \otimes L_{1} \otimes L_{2} -L_{1} \otimes L_{1} \otimes L_{4} +L_{1} \otimes L_{2} \otimes L_{1} +L_{1} \otimes L_{2} \otimes L_{2}\\
 & -L_{1} \otimes L_{4} \otimes L_{1} -L_{1} \otimes L_{4} \otimes L_{4} +L_{2} \otimes L_{1} \otimes L_{1} +L_{2} \otimes L_{1} \otimes L_{2}\\
 & +L_{2} \otimes L_{2} \otimes L_{1} -L_{2} \otimes L_{2} \otimes L_{3} -L_{2} \otimes L_{3} \otimes L_{2} -L_{2} \otimes L_{3} \otimes L_{3}\\
 & -L_{5} \otimes L_{1} \otimes L_{1} -L_{5} \otimes L_{1} \otimes L_{4} -L_{5} \otimes L_{2} \otimes L_{2} -L_{5} \otimes L_{2} \otimes L_{3}\\
 & +L_{5} \otimes L_{3} \otimes L_{2} +L_{5} \otimes L_{3} \otimes L_{3} +L_{5} \otimes L_{4} \otimes L_{1} +L_{5} \otimes L_{4} \otimes L_{4}\\
 & -2L_{5} \otimes L_{5} \otimes L_{1} -2L_{5} \otimes L_{5} \otimes L_{2} +2L_{5} \otimes L_{5} \otimes L_{3} +2L_{5} \otimes L_{5} \otimes L_{4},
\end{aligned}
\end{equation}
where the letters are defined as $\{L_1, L_2,\ldots,L_5\} = \{X_1+Y,X_2+Y,X_1-Y,X_2-Y,X_1+X_2\}$.

For the coaction calculated in the main text, we perform $\epsilon $-expand, using \texttt{HypExp}~\cite{Huber:2007dx}, for every entry of the tensor product. Below, we checked they are consistent at each order of $\epsilon $.

\subsection{Cancellation at \texorpdfstring{$\mathcal{O}(\epsilon^{-2})$}{O(epsilon minus 2)} and \texorpdfstring{$\mathcal{O}(\epsilon^{-1})$}{O(epsilon minus 1)}}
So we firstly check the cancellation at $\mathcal{O} (\epsilon ^{-2} )$ and $\mathcal{O} (\epsilon ^{-1} )$. At $\mathcal{O} (\epsilon ^{-2} )$, we have
\begin{equation}
\relax[ \Delta J_{\{1,2,3\}}]| _{\epsilon ^{-2}} =\sum _{X=\{1,2\},\{1,3\},\{2,3\}} J_{X}| _{\epsilon ^{-2}} \otimes \mathcal{C}_{X} J_{\{1,2,3\}}| _{\epsilon ^{0}} ,
\end{equation}
where $X=\{3\}$ term is omitted since it starts at $\mathcal{O} (\epsilon ^{-1} )$, and the left and right entry are all of weight-0, {\it i.e.} rational numbers. And  $\relax[ \Delta J_{\{1,2,3\}}]| _{\epsilon ^{-2}} =0$ is easily checked by linearity of the tensor product.

At $\mathcal{O} (\epsilon ^{-1} )$, we have 
\begin{equation}
\begin{aligned}
\relax[  \Delta J_{\{1,2,3\}}]| _{\epsilon ^{-1}} & =\sum _{X=\{1,2\},\{1,3\},\{2,3\}} J_{X}| _{\epsilon ^{-2}} \otimes \mathcal{C}_{X} J_{\{1,2,3\}}| _{\epsilon ^{1}}\\
 & +\sum _{X=\{1,2\},\{1,3\},\{2,3\}} J_{X}| _{\epsilon ^{-1}} \otimes \mathcal{C}_{X} J_{\{1,2,3\}}| _{\epsilon ^{0}}\\
 & + J_{\{3\}}| _{\epsilon ^{-2}} \otimes \mathcal{C}_{\{3\}} J_{\{1,2,3\}}| _{\epsilon ^{0}} ,
\end{aligned}
\end{equation}
which contains $\Delta _{0,1}$ and $\Delta _{1,0}$ component. Considering the second entry of coaction is defined modulo $\pi $, we can freely change $\log (x)$ to $\log (-x)$, which enables us to verify $[ \Delta J_{\{1,2,3\}}]| _{\epsilon ^{-1}} =0$.

\subsection{Verification at \texorpdfstring{$\mathcal{O}(\epsilon^{0})$}{O(epsilon 0)}}
At $\mathcal{O} (\epsilon ^{0} )$, we only need to consider the following order of $\epsilon $ of the left/right entry of the coaction and it directly gives the result of symbol:
\begin{equation}
\Delta _{1,1}[ J_{\{1,2,3\}}| _{\epsilon ^{0}}] =\sum _{X=\{1,2\},\{1,3\},\{2,3\}} J_{X}| _{\epsilon ^{-1}} \otimes \mathcal{C}_{X} J_{\{1,2,3\}}| _{\epsilon ^{1}} + J_{\{3\}}| _{\epsilon ^{-1}} \otimes \mathcal{C}_{\{3\}} J_{\{1,2,3\}}| _{\epsilon ^{0}} .
\end{equation}
We expand the every $J_{X}$ and $\mathcal{C}_{X} J_{\{1,2,3\}}$, which is
\begin{equation}
\begin{array}{ l }
J_{\{1,2\}} |_{\epsilon ^{-1}} \otimes \mathcal{C}_{\{1,2\}} J_{\{1,2,3\}} |_{\epsilon ^{1}} =\log (L_{1} L_{2} )\otimes \log (L_{1} L_{2} )\\
J_{\{1,3\}} |_{\epsilon ^{-1}} \otimes \mathcal{C}_{\{1,3\}} J_{\{1,2,3\}} |_{\epsilon ^{1}} =\log L_{1} \otimes \log (L_{1} L_{4} )\\
J_{\{2,3\}} |_{\epsilon ^{-1}} \otimes \mathcal{C}_{\{2,3\}} J_{\{1,2,3\}} |_{\epsilon ^{1}} =\log L_{2} \otimes \log (L_{2} L_{3} )\\
J_{\{3\}} |_{\epsilon ^{-1}} \otimes \mathcal{C}_{\{3\}} J_{\{1,2,3\}} |_{\epsilon ^{0}} =-\log L_{5} \otimes \log\left(\frac{L_{1} L_{2}}{L_{3} L_{4}}\right).
\end{array}
\end{equation}
Putting them all together, we verify the consistency with $\mathcal{O} (\epsilon ^{0} )$ in previous result.

\subsection{Verification at \texorpdfstring{$\mathcal{O}(\epsilon^{1})$}{O(epsilon 1)} and coassociativity}
At $\mathcal{O}(\epsilon^{1})$, the result has transcendental weight three. Accordingly, the coaction contains two relevant components, namely $\Delta_{1,2}$ and $\Delta_{2,1}$.
We explicitly write the two coactions as follows
\begin{equation}
\begin{aligned}
\Delta _{1,2}[ J_{\{1,2,3\}}| _{\epsilon ^{1}}] =\sum _{X=\{1,2\},\{1,3\},\{2,3\}} J_{X}| _{\epsilon ^{-1}} \otimes \mathcal{C}_{X} J_{\{1,2,3\}}| _{\epsilon ^{2}} + J_{\{3\}}| _{\epsilon ^{-1}} \otimes \mathcal{C}_{\{3\}} J_{\{1,2,3\}}| _{\epsilon ^{1}} ,\\
\Delta _{2,1}[ J_{\{1,2,3\}}| _{\epsilon ^{1}}] =\sum _{X=\{1,2\},\{1,3\},\{2,3\}} J_{X}| _{\epsilon ^{0}} \otimes \mathcal{C}_{X} J_{\{1,2,3\}}| _{\epsilon ^{1}} + J_{\{3\}}| _{\epsilon ^{0}} \otimes \mathcal{C}_{\{3\}} J_{\{1,2,3\}}| _{\epsilon ^{0}} ,
\end{aligned}
\end{equation}
where
\begin{equation}
\begin{aligned}
J_{\{1,2\}} |_{\epsilon ^{-1}} \otimes \mathcal{C}_{\{1,2\}} J_{\{1,2,3\}} |_{\epsilon ^{2}} & =\log (L_{1} L_{2} )\otimes \frac{1}{2}\log^{2} (L_{1} L_{2} )\\
J_{\{1,3\}} |_{\epsilon ^{-1}} \otimes \mathcal{C}_{\{1,3\}} J_{\{1,2,3\}} |_{\epsilon ^{2}} & =\log L_{1} \otimes \frac{1}{2}\log^{2} (L_{1} L_{4} )\\
J_{\{2,3\}} |_{\epsilon ^{-1}} \otimes \mathcal{C}_{\{2,3\}} J_{\{1,2,3\}} |_{\epsilon ^{2}} & =\log L_{2} \otimes \frac{1}{2}\log^{2} (L_{2} L_{3} )\\
J_{\{3\}} |_{\epsilon ^{-1}} \otimes \mathcal{C}_{\{3\}} J_{\{1,2,3\}} |_{\epsilon ^{1}} & =-\log (L_{5} )\otimes \Bigl[ -2\text{Li}_{2}\left(\frac{L_{5}}{L_{2}}\right) +2\text{Li}_{2}\left(\frac{L_{5}}{L_{4}}\right)\\
 & \hspace{-10em}+\frac{1}{2}\left(\log\left( -\frac{L_{1}}{L_{4}}\right) -\log\left( -\frac{L_{3}}{L_{2}}\right)\right)\left(\log\left( -\frac{L_{3}}{L_{2}}\right) +\log\left( -\frac{L_{1}}{L_{4}}\right) +4\log (-L_{5} )\right)\Bigr],
\end{aligned}
\end{equation}
and
\begin{equation}
\begin{aligned}
J_{\{1,2\}} |_{\epsilon ^{0}} \otimes \mathcal{C}_{\{1,2\}} J_{\{1,2,3\}} |_{\epsilon ^{1}} & =\left(\frac{1}{2}\log^{2} (L_{1} L_{2} )+\frac{\pi ^{2}}{3}\right) \otimes \log (L_{1} L_{2} )\\
J_{\{1,3\}} |_{\epsilon ^{0}} \otimes \mathcal{C}_{\{1,3\}} J_{\{1,2,3\}} |_{\epsilon ^{1}} & =\Bigl[\text{Li}_{2}\left(\frac{L_{5}}{L_{4}}\right) +\frac{1}{12}\left( 6\log^{2} (L_{1} )-6\log^{2} (L_{4} )+12\log L_{4}\log L_{1} -\pi ^{2}\right)\\
 & +\log\left( -\frac{L_{1}}{L_{4}}\right)\log\left(\frac{L_{5}}{L_{4}}\right)\Bigr] \otimes \log (L_{1} L_{4} )\\
J_{\{2,3\}} |_{\epsilon ^{0}} \otimes \mathcal{C}_{\{2,3\}} J_{\{1,2,3\}} |_{\epsilon ^{1}} & =\Bigl[ -\text{Li}_{2}\left(\frac{L_{5}}{L_{2}}\right) +\log^{2} (L_{2} )+\log\left( -\frac{L_{3}}{L_{2}}\right)\log\left(\frac{L_{2}}{L_{5}}\right) +\frac{5\pi ^{2}}{12}\Bigr] \otimes \log (L_{2} L_{3} )\\
J_{\{3\}} |_{\epsilon ^{0}} \otimes \mathcal{C}_{\{3\}} J_{\{1,2,3\}} |_{\epsilon ^{0}} & =\left( -\log^{2} L_{5} -\frac{\pi ^{2}}{4}\right) \otimes \log\left(\frac{L_{1} L_{2}}{L_{3} L_{4}}\right).
\end{aligned}
\end{equation}
The property of coassociatity requires the symbol calculated from different coaction $\Delta _{1,2}$ and $\Delta _{2,1}$ to be the same:
\begin{equation}
\Delta _{1,1,1} =\left(\text{id} \otimes \Delta _{1,1}\right) \Delta _{1,2} =\left( \Delta _{1,1} \otimes \text{id}\right) \Delta _{2,1} ,
\end{equation}
which can be easily verified and has been checked to be the same as $\mathcal{O} (\epsilon ^{1} )$ of Eq.~\eqref{eq:2sitew3}.

\bibliographystyle{JHEP}
\bibliography{inspire.bib}

\end{document}